\documentclass[twocolumn,showpacs,preprintnumbers,amsmath,amssymb]{revtex4}

\usepackage{graphicx,color}
\usepackage{dcolumn}
\usepackage{bm}

\newcommand{\bra}[1]{\langle #1|}
\newcommand{\ket}[1]{|#1\rangle}

\def\>{\rangle}
\def\<{\langle}

\def\tr{ \mbox{tr} }

\begin{document}


\title{Maximally and minimally correlated states attainable within a closed evolving system}

\author{Sania Jevtic, David Jennings and Terry Rudolph}
 \affiliation{Controlled Quantum Dynamics Theory, Department of Physics, Imperial College London, London SW7 2AZ}

\date{October 12, 2011}

\begin{abstract}
The amount of correlation attainable between the components of a quantum system is constrained if the system is closed. We provide some examples, largely from the field of quantum thermodynamics, where knowing the maximal possible variation in correlations is useful. The optimization problem it raises requires us to search for the maximally and minimally correlated states on a unitary orbit, with and without energy conservation. This is fully solvable for the smallest system of two qubits. For larger systems, the problem is reduced to a manageable, classical optimization. 
\end{abstract}

\pacs{03.65.Ta,  03.67.Mn,  05.70.Ln}
\maketitle

The idealized notion of a closed system is central to both classical and quantum mechanics, across scales from the microscopic to the universe itself. Here, we concern ourselves with the quantum mechanical version of a fundamental question: In the interactions between the constituent components of a closed system, to what extent does the closure of the system constrain the correlations attainable?

We focus on the simplest case, where we divide the closed system into two parts and the correlations between these are quantified by the mutual information. For a given bipartite state of the system we therefore seek the two extremal (minimally and maximally) correlated states under all evolution that does not change the total entropy. We will also consider the case of  evolution that obeys the additional restriction of energy conservation, either in a weak sense (the expected energy stays constant) or a strong sense (the interaction commutes with the free Hamiltonians of the two subsystems).

We find that the answer to these problems, particularly for the case of the minimal attainable correlation, has a surprisingly rich mathematical structure. Because of the foundational nature of this result it can be applied to a range of problems. Before turning to our technical results, we present in some detail three such examples from the field of quantum thermodynamics.

\emph{Example 1: Environmentally friendly work extraction from a Szilard Engine.} Our first example concerns a Szilard engine immersed in a thermal bath at temperature $T$ using correlated particles from which to extract work. The engine admits individual subsystems, one at a time, to ``burn as fuel''. We consider the case of two quantum subsystems, described by a bipartite mixed state $\rho$. For such fuel reserves, we can extract \cite{OppHorodeckis} from each subsystem at most an amount of work $W_\mu=kT(\log d_\mu - S(\rho_\mu))$ where $d_\mu$ is the dimension of subsystem $\mu \in \{A,B\}$, $\rho_\mu$ is its state, and $S(\rho_\mu) = -\tr(\rho_\mu \log \rho_\mu)$ is its von Neumann entropy. The goal is to increase the total work extracted from the pair of systems:
\[
W = W_A + W_B = kT(\log d_A d_B - S(\rho_A)-S(\rho_B)).
\]
To do so, before the systems are fed into the engine they are sent into a refinery whose purpose is to ``purify'' $\rho_A$ and $\rho_B$ so as to reduce $S(\rho_A)+S(\rho_B)$. More accurately, the refinery tries to localize existing purity in the composite fuel state. Such a purification scheme has been considered before under the restriction of local operations and classical communication (LOCC) processes   \cite{OppHorodeckis,AlickiHorodeckis},
however, here we work in a broader context and permit a global operation on the composite fuel state $\rho$, but crucially we impose the restriction that the refining process, which takes $\rho$ to $\rho'$, must be ``environmentally friendly'' in the sense that all measures of purity, such as the von Neumann entropy or $\tr[\rho^2]$, remain constant \footnote{In particular, the requirement of constant total entropy during the purification is particularly natural given the subtle entropic counting that needs to be performed in any analysis of Szilard engines.}. As a result, we are forced into taking the refining process to be a global unitary operation on the full reserve of fuel.

The extra mechanical work obtained through the refining process is
$W^{\rm extra} = -kT(\Delta S_A+\Delta S_B) = -kT \Delta I$
where $\Delta S_\mu = S(\rho'_\mu) - S(\rho_\mu)$, $\Delta I = I(\rho') - I(\rho)$, and we have introduced the quantum mutual information (QMI) $I(\rho) = S(\rho_A) + S(\rho_B) - S(\rho) \geq 0$, which is the natural measure of correlations. If $A$ and $B$ are initially uncorrelated, the QMI is at its minimum and cannot be reduced; $W^{\rm extra} = 0$. However, if correlations are initially present in $\rho$, it is possible to obtain $|W^{\rm extra}| > 0$: a natural challenge is to find the maximum $|W^{\rm extra}|$ for a given initial state fuel reserve $\rho$, in other words, to determine the largest attainable $|\Delta I|$ under the environmentally friendly constraint. Generically, it is impossible to fully decorrelate the state, and the optimal refinement process reduces to the broader problem under consideration in this article.

\emph{Example 2: Anomalous heat flow in the presence of correlations.}
It is known for two subsystems of a closed system, each initially in thermal states, that the traditional thermodynamic flow of heat from hot to cold can be distorted by the presence of correlations \cite{Partovi,JennRud}. Indeed, with sufficiently strong correlations, a substantial amount of heat can be made to flow anomalously from the colder to the hotter system. What are the limitations on this process? Again, let $\rho$ be the initial joint state of the two systems, $\mu \in \{A,B\}$. By assumption, each subsystem is initially in a thermal (Gibbs) state $\rho_\mu = \rho^{\rm th}_\mu=e^{-\beta_\mu H_\mu} / Z_\mu$ at temperature $\beta_\mu^{-1}=kT_\mu$, where $Z_\mu = \tr(e^{-H_\mu/T_\mu})$, is the partition function. The subsystems interact, either by switching on a known controlled interaction for some finite time or by a scattering process, and the composite state $\rho$ evolves to a final state $\rho'$, which has local states $\rho'_A,\rho'_B$.

The free energy functional $F_{H,T} [\rho] := \tr(\rho H) - kT S(\rho)$ is obtained from the relative entropy function with respect to the Gibbs state and is defined over the full state space. It is minimized by the thermal state $e^{-\beta H}/\tr(e^{-\beta H})$, $\beta^{-1} = kT$, and its value coincides with the usual thermodynamic free energy. Thus each subsystem satisfies the inequality $F_{H_\mu,T_\mu} [\rho'_\mu] - F_{H_\mu,T_\mu} [\rho_\mu] \geq 0$ \textit{for any state} $\rho'_\mu$ (originating from the positivity of the relative entropy) which when added together yield
\begin{align}
\label{betaQS} \beta_A Q_A + \beta_B Q_B \geq \Delta S_A + \Delta S_B,
\end{align}
where $Q_\mu = \tr (\rho'_\mu H_\mu) - \tr (\rho_\mu H_\mu)$ is the heat \footnote{Heat is usually defined as $Q_\mu = \tr (\rho'_\mu H_\mu) - \tr (\rho_\mu H_\mu)$ (see for instance A. Peres \textit{Quantum Theory: Concepts and Methods}, Springer (1995)) but it is assumed that the initial and final states are both diagonal in $H_\mu$. In the heat flow model this is true for the initial state, however for the final state $\left[\rho'_{\mu}, H_\mu\right] \neq 0$ is permitted. This is not fundamentally a problem when we remember that the only property being measured is the changes in the observables $H_A, H_B$, i.e. the changes in local energies, and this is called ``heat'' because the local entropies vary and the energies exchanged between A and B are assumed inaccessible for external work. In our model the experimenter is not required to know the initial correlations nor the interaction Hamiltonian so we need not appeal to a generalized notion of work and heat such as that proposed in H. Weimer, M. J. Henrich, F. Rempp, H. Schr\"{o}der, and G. Mahler EPL \textbf{83} 30008 (2008).} into system $\mu$.
Note that this inequality only demands that an initial temperature be defined, and no further restrictions on $\rho'_\mu$ are needed at this stage. Under the closed system constraint of constant total entropy \textit{and} constant energy, $Q_A + Q_B = 0$, we can write (\ref{betaQS}) as
\begin{align}
\label{QTI} Q_A \left(\frac{1}{kT_A} - \frac{1}{kT_B}\right) \geq \Delta I.
\end{align}
This inequality provides directionality for any energy conserving process. It relies on local initial properties but also depends on non-local correlations. Any initial correlations, up to the constraint of thermal marginals, are permitted and the bound is independent of any assumptions on interaction strength, in contrast to several previous considerations of the thermodynamics of open quantum systems where weak coupling between the system and the bath is required \cite{WeakStrCoup1,WeakStrCoup2,CorrEntropy}. We are interested in the evolution of a closed system which in itself displays thermodynamic behaviour.
In standard thermodynamics it is assumed that the interacting systems are initially uncorrelated, rendering the entropy as additive: $\rho = \rho_{A}\otimes\rho_{B}$ and thus $I(\rho) = 0$. As the interaction cannot decorrelate $A$ and $B$ any further $I(\rho') \geq I(\rho)$ and it follows that the left hand side of equation \ref{QTI} must be positive.
This means that when $T_A \leq T_B$ it must be the case that $Q_A \geq 0$, and heat flows in the standard manner, from hot to cold.

In general, however, systems $A$ and $B$ could initially possess correlations \footnote{It is reasonable to assume that A and B are ``locally thermal'', that is, when one is restricted to doing only local operations on them, they are indistinguishable from uncorrelated thermal states with matching local Hamiltonians and temperatures. Although they are locally thermal they can still be correlated.}, in which case the interaction could lower the QMI. If $\Delta I < 0$ then there is no longer an absolute restriction on the direction of heat flow and for a suitably chosen interaction we will deterministically observe heat being transferred from the colder to the hotter body. We call this anomalous heat flow (AHF). Even though the local entropies have decreased and negative heat flow has occurred, after the local measurement of the individual energies the system is left uncorrelated and thus one cannot cause heat to flow from cold to hot in a \textit{cyclic} process, thus saving the second law. In this sense correlations are a resource.

To observe a large AHF, the initial state of the system would have to be very correlated, possibly entangled. Indeed, the AHF constitutes a discriminating feature between quantum and classical thermodynamics, and may be used as an operational indicator of entanglement \cite{JennRud} that does not require knowledge of the joint initial state of the two systems!
This is easily seen, since the QMI over separable states is bounded from above by $\log (\min\{d_A,d_B\})$, while for the full quantum state space the bound is twice this. Therefore when $\Delta I > \log (\min\{d_A,d_B\})$ the initial state $\rho$ must be non-separable, and in turn, any transfer of heat from the colder to the hotter body of an amount greater than $\frac{\log (\min\{d_A,d_B\})}{|\beta_A - \beta_B|}$ indicates the presence of entanglement \cite{JennRud}.

Keeping in mind the additional constraint of equal energies for $\rho$ and $\rho'$ included in this example, the quantity of AHF possible in a closed system is bounded by the largest $\Delta I$ that can be obtained reversibly. Once again, the determination of such a fundamental limitation reduces to our general problem.

\emph{Example 3: Partovi/Peres collision model of equilibration.}
In Ref. \cite{PartoviOld} Partovi proposed a collision model of equilibration, later simplified by Peres \cite{Peres}. Two ingredients are required in the collision process: firstly an increase in the local entropies, which is achieved by interacting two initially uncorrelated quantum systems via a (strongly) energy conserving unitary, and secondly irreversibility, causing a growth of the total entropy of the system. In the model the latter is enforced by assuming that the two systems decorrelate after interacting. One full collision can be written as $\rho = \rho_A \otimes \rho_B \rightarrow \rho' = U \rho U^{\dag} \rightarrow \rho'_A \otimes \rho'_B$, with $S(\rho'_A) + S(\rho'_B) \geq S(\rho_A) + S(\rho_B)$. This process is reiterated, and it can be shown the systems reach a stationary state of equal temperature.

The second requirement of complete decorrelation to a product state is very stringent - given that physical systems typically dephase (i.e. off diagonal ``coherences'' of the density matrix decay) much more rapidly than they completely decorrelate. A natural question therefore is whether the systems can retain some minimal amount of correlation and still reach equilibrium. Part of the solution to examples 1 and 2 is finding the state which has the minimum QMI on a unitary orbit: when the two interacting particles are qubits, we can use this result to show that, after the unitary part of the collision, if the qubits dephase to this minimally correlated state (which is not a product state) then equilibration is still achieved \cite{NextOne}.


\emph{Overview of the general solution:}
Given an $N = d_A d_B$-dimensional bipartite state $\rho$ with spectrum $\Lambda = \{\lambda_i \}$; our goal is to find $\rho_{\min}$ ($\rho_{\max}$) defined, modulo local unitary transformations, as the state for which $I$ is minimal  (maximal) over the unitary orbit \cite{ModiUO}, $\mathcal{O} = \{\tau : \tau = U \rho U^{\dag}\}$, for all unitaries $U$ of dimension $N$. For simplicity we do not demand energy conservation for now but revisit it later when we consider a two qubit system.

Finding the maximally correlated state is hard classically \cite{NextOne} but fairly straight forward over the space of quantum states. We can always find a unitary that transforms a state to
\begin{equation}
\label{eqrhomax}
\rho_{\max} = \sum_{i=1}^{N'} \lambda_i \ket{\Phi_i}\bra{\Phi_i},
\end{equation}
where $\{ \ket{\Phi_i} \}$ is any generalized Bell state basis \cite{GBS} with $N' = (\min\{d_A,d_B\})^2$, obtained from the Schmidt decomposition. Since $\tr_A(\ket{\Phi_i}\bra{\Phi_i}) \propto \mathbb{I}_B $ for all $i$ we deduce that also $\tr_A(\rho_{\max}) \propto \mathbb{I}_B$ and in turn $I(\rho_{\max}) = 2\log(\min\{d_A,d_B\}) - H(\Lambda)$. This is the maximum attainable value of the QMI over all state space, with a reduction by the amount $H(\Lambda) = -\sum_i \lambda_i \log \lambda_i$, the Shannon entropy, because of the restriction to a unitary orbit.

Finding the minimally correlated state is considerably harder: because the total spectrum of the state is fixed, given an initial state $\rho$, there does not always exist a unitary transformation that can decorrelate its subsystems. Hence $I(\rho_{\min}) \geq 0$ in general and, unlike $\rho_{\max}$, the minimum sum of the local entropies depends on $\Lambda$. The challenge is to optimize over the set of reduced states compatible with a composite system having a fixed spectrum $\Lambda$. Finding the set of allowed such reduced states is the highly nontrivial ``quantum marginal problem'' \cite{Klyachko,Matthias,Bravyi}.

The initial difficulty is that the optimization problem is not convex. There does not even appear to be a simple argument that the minimally correlated state should be separable, although intuitively it seems reasonable that this should be the case.

In fact we are able to prove something stronger: the minimum of the quantum mutual information $I(\rho)$ over the unitary orbit is attained for a \emph{classically} correlated state
\begin{align}
\label{eqrhomingen} \rho_{\min} = \sum_{j,k} \lambda_{jk} \ket{e_j}\bra{e_j}\otimes\ket{f_k}\bra{f_k},
\end{align}
where $\lambda_{jk}$, $j =1,..,d_A,k=1..d_B$ is a reindexing of $\lambda_i$
and $\{\ket{e_j}\}$, $\{\ket{f_k}\}$ are orthonormal basis states for systems $A$ and $B$. That is, the minimum of the QMI over the unitary orbit
equals $H(\sum_j\Pi(\lambda_{jk})) + H(\sum_k\Pi(\lambda_{jk})) - H(\Lambda)$, where the first two terms are the Shannon entropies 
for the marginal of some permutation ($\Pi$) of the eigenvalues $\lambda_{jk}$.

To prove this, we consider the function $G[\sigma_{A},\sigma_{B}] = S[\sigma_A] + S[\sigma_B]$ defined over the convex hull $\mathcal{C}$ of the unitary orbit $\mathcal{O}$ of $\rho$. The states in $\mathcal{C}$ take the form $\sigma = \sum_i p_i U_i \rho U_i^\dagger$, with $\sum p_i = 1$, $p_i \geq 0$ and $\sigma_A, \sigma_B$ are the reduced states of $\sigma$. We then look for the minima of this function $G$. If these happen to occur on the unitary orbit, where $S(\rho)$ in constant, then it will also give us the minima of $I$ over $\mathcal{O}$.

Writing the eigenvalues as components of vectors, $\boldsymbol{\nu} = \mathrm{spec}(\sigma)$ and $\boldsymbol{\lambda}=\mathrm{spec}(\rho)$, it can be shown that the reduced states of any $\sigma$ (which include the unitary orbit states) have eigenvalues that are marginals of a probability distribution obeying the majorization relation $\boldsymbol{\nu} \prec \boldsymbol{\lambda}$ \cite{Bravyi}. Note that all $\boldsymbol{\nu}$ satisfying this relation form a convex set $\mathcal{P(\boldsymbol{\lambda})}$. $G$ can be shown to be concave on the set $\mathcal{P(\boldsymbol{\lambda})}$, and so its minima occur at the extremal points. These extrema are permutations of the components of $\boldsymbol{\lambda}$, whose corresponding states lie on the unitary orbit, and so the minimum QMI occurs at a permutation of the $\{\lambda_i\}$ \footnote{The state of minimal correlations $\rho_{\min}$ is not unique due to symmetries of the QMI which are local unitary operations and a swap of A and B states.}.

However, knowing that the state is classical is not the full solution to the problem. Consider a state with $\mathrm{spec}(\rho)=(1/2,1/2,0,0)$ - the two classical states of the form (\ref{eqrhomingen}) $(\ket{00}\bra{00}+\ket{11}\bra{11})/2$ and $(\ket{00}\bra{00}+\ket{01}\bra{01})/2$ have the correct spectrum but the former is correlated while the latter is not. So the QMI depends on the ordering of the eigenvalues in $\rho_{\min}$.

There are $N!$ different permutations of $\lambda_i$ to consider, however it is possible \cite{NextOne} to reduce  this number down to an irreducible set of Young Tableaux \cite{WuKiTung} in which the minimally correlated state will be found. For the simplest case of $d_A=d_B=2$ the set has a unique element, which can be compactly represented
\begin{equation}\label{2qubittableux}
\left[\nu_{ij}\right]=
\left[
  \begin{array}{cc}
    \lambda_1 & \lambda_2 \\
    \lambda_3 & \lambda_4 \\
  \end{array}
\right].
\end{equation}
Here the eigenvalues $\lambda_i$ are in non-increasing order, and row $j$ column $k$ corresponds to the re-indexing element $\lambda_{jk}$ of Eq. (\ref{eqrhomingen}) above. For $d_A=2,d_B=3$ the full set of permutations has $(d_Ad_B)!=720$ elements, however our analysis \cite{NextOne} reduces this to just 5 tableaux:
\[
\left[
  \begin{array}{ccc}
    \lambda_1 & \lambda_2 & \lambda_3 \\
    \lambda_4 & \lambda_5 & \lambda_6 \\
  \end{array}
\right],
\left[
  \begin{array}{ccc}
    \lambda_1 & \lambda_2 & \lambda_4 \\
    \lambda_3 & \lambda_5 & \lambda_6 \\
  \end{array}
\right],
\left[
  \begin{array}{ccc}
    \lambda_1 & \lambda_2 & \lambda_5 \\
    \lambda_3 & \lambda_4 & \lambda_6 \\
  \end{array}
\right],
\]
\[
\left[
  \begin{array}{ccc}
    \lambda_1 & \lambda_3 & \lambda_4 \\
    \lambda_2 & \lambda_5 & \lambda_6 \\
  \end{array}
\right],
\left[
  \begin{array}{ccc}
    \lambda_1 & \lambda_3 & \lambda_5 \\
    \lambda_2 & \lambda_4 & \lambda_6 \\
  \end{array}
\right].
\]
For the case of two qutrits there are 21 tableaux to consider, for two 4-dimensional systems the irreducible set has approximately 12000 elements. Clearly it would be desirable to have an efficient algorithmic procedure to identify the element on the irreducible set on which the minimum is attained, but it is currently not clear if one exists.

\emph{The primitive case of two qubits:}
As an illustrative example we consider two qubits in which case the preceding discussion shows that the minimal QMI on a unitary orbit has a value of
\[
I(\rho_{\min})= H(\lambda_1+\lambda_2) + H(\lambda_1+\lambda_3) - H(\Lambda),
\]
where $\lambda_1\ge \lambda_2\ge \lambda_3\ge \lambda_4$ and we have used the notation for the binary Shannon entropy 
$H(x) = -x\log x - (1-x)\log (1-x)$ in the first two terms. 
Therefore the maximum that the QMI can change by for a two qubit system undergoing a global unitary transformation is
\begin{align}
\label{deltaI}\Delta I^{\max}_U = 2 - H(\lambda_1+\lambda_2) - H(\lambda_1+\lambda_3).
\end{align}

Considerable insight into this case can be gained by doing the optimization more explicitly. This is possible because the quantum marginal problem for a composite system of two qubits has been solved \cite{Bravyi} and the results are readily applied to our situation. Examining this also allows us to include the constant energy constraint.

\begin{figure}
\includegraphics[width=3.2in]{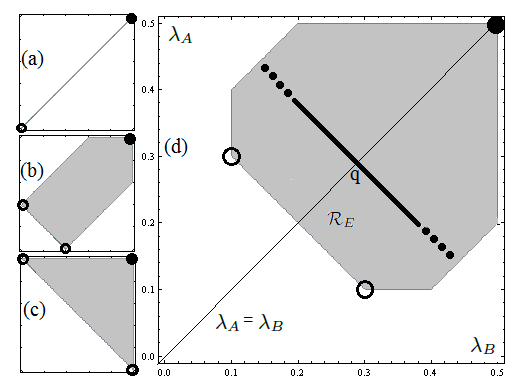}
\caption{Regions $\mathcal{R}$ (shaded) of allowed  $\lambda_A$ ($y$-axis), $\lambda_B$ ($x$-axis) when the joint state of two qubits, with spectrum $\Lambda = \{\lambda_i\}_{i=1}^4$, has various ranks: $\Lambda =$ (a) \{1,0,0,0\}, (b) \{0.8,0.2,0,0\}, (c) \{0.5,0.5,0,0\}, (d) \{0.6,0.3,0.1,0\}. $\lambda_A, \lambda_B \in [0,\frac{1}{2}]$.
For each spectrum, the hollow circles correspond to $\rho_{\min}$, the large filled ones to $\rho_{\max}$.
In (d), a state with energy $E$ defines the set $\mathcal{R}_E$ of states which could also have energy $E$. It is bounded from ``above'' by the solid-and-dotted line, on which the state itself is situated. The maximally correlated state in $\mathcal{R}_E$ is at q.}
\label{LALB_ranks}
\end{figure}

Let us denote the two eigenvalues of the reduced state $\rho_{\mu}$ as $\lambda_{\mu},\, 1-\lambda_{\mu}$ where $\lambda_{\mu} \leq \frac{1}{2}$, $\mu \in \{ A,B \}$. There is a set of inequalities that constrain the spectra of these marginals, given $\Lambda$, to a set $\mathcal{R}$ \footnote{Let the eigenvalues $\lambda_i$ in spectrum $\Lambda$ of the joint state $\rho$ be arranged in non-increasing order $\lambda_1 \geq \lambda_2 \geq \lambda_3 \geq \lambda_4 $ and denote the two eigenvalues of the reduced state $\rho_\mu$ as $\lambda_\mu,\, 1-\lambda_\mu$ where $0\leq \lambda_\mu \leq \frac{1}{2}$, $\mu \in \{ A,B \}$. Then $\rho_A, \rho_B$ are valid marginals of a state in $\mathcal{O}$ if and only if their eigenvalues satisfy the following inequalities:
$\lambda_A \geq \lambda_3 + \lambda_4, \,
\lambda_B \geq \lambda_3 + \lambda_4, \,
\lambda_A + \lambda_B \geq \lambda_2 + \lambda_3 + 2\lambda_4, \,
| \lambda_A - \lambda_B | \leq \operatorname{min}\{ \lambda_1-\lambda_3 , \lambda_2- \lambda_4 \}.$
These inequalities define the set $\mathcal{R}$ and this result is taken from S. Bravyi, Quant. Inf. and Comp. \textbf{4}, 012 (2008)}.  Figure \ref{LALB_ranks} depicts the shape of the set $\mathcal{R}$ (shaded) that $\lambda_A, \lambda_B$ occupy and gives some representative examples of how it varies according to the rank of $\rho$.

\emph{Two qubit correlations with energy conservation:}
Example 2 above sought the maximal change in the QMI for a bipartite system in a state $\rho$ undergoing unitary evolution to a new state $\rho'$ and constrained to energy conservation $\tr(\rho H) = \tr(\rho' H) :=E$, where $H = H_A +H_B$ is the sum of the original local Hamiltonians. The reduced states of $\rho'$ are allowed to be non-diagonal in $H_A, H_B$. This divides the set $\mathcal{R}$ of allowed reduced states into two regions: ones that could have energy $E$, forming the set $\mathcal{R}_E\subseteq \mathcal{R}$, and ones that could not. $\mathcal{R}_E$ defines an ``energy-conserving region''. For simplicity, let us pick $H_A = H_B = \ket{1}\bra{1}$, so the energy spacing of $H_A$ equals that of $H_B$. The region $\mathcal{R}_E$ is shown in figure \ref{LALB_ranks} (d). It is shown in Ref. \cite{NextOne} that
the maximal variation of correlations for a two qubit state undergoing an energy conserving unitary transformation is found to be
$$ \Delta I_E^{\max} = 2H\left(\frac{E}{2}\right) - H(\lambda_1 + \lambda_2) - H(\lambda_1 + \lambda_3),$$
where the maximally and minimally correlated states in $\mathcal{R}_E$ are also shown in figure \ref{LALB_ranks}.

An interesting observation is that the point q in the figure does not uniquely define a joint state (even up to local unitaries). It can be the case that a strong energy conserving unitary acting on one state at $q$ transform it only along the solid portion of the line however it evolves another along the full solid-and-dotted line. This is because these two states have different types of correlation even though they have the same QMI.
The details for this appear in Ref. \cite{NextOne}. In any case the set of states reached in $\mathcal{R}_E$ is restricted to the line for strong energy conserving unitary evolutions. These states have minimal variance for energy measurements.
Weak energy conserving unitaries can transform the initial state to all other points in $\mathcal{R}_E$, which involve intrinsically quantum fluctuations via superpositions.

\emph{Conclusions:}
In this Letter, we have analysed the abstract problem of how correlations vary along unitary orbits for isolated quantum systems, an intricate mathematical task that reveals a complex relationship between the mutual information and the ordering of a bipartite probability distribution. The results of this find application in different thermodynamic scenarios such as equilibration, heat exchange and localisation of free energy. Our work can be extended to understanding the correlation structure of more complicated processes, such as a quantum channel consisting of $k$ unitaries each applied with some probability $p_k$ to the bipartite state. It would also be of interest to explore connections between our work and the recent papers \cite{ResourceTD1,ResourceTD2} on the resource theory of quantum thermodynamics.


We wish to acknowledge R. Spekkens and J. Anders for their useful comments. This work was supported by EPSRC. 

\bibliography{twoqubitref}

\end{document}